# Immersive Technologies in Virtual Companions: A Systematic Literature Review

Ziaullah Momand, Jonathan H. Chan, Pornchai Mongkolnam

Ziaullah.momand1@kmutt.ac.th, jonathan@sit.kmutt.ac.th, pornchai@sit.kmutt.ac.th

School of Information Technology, King Mongkut's University of Technology Thonburi, 126 Pracha Uthit Rd, Bang Mot, Thung Khru, Bangkok 10140, Thailand

## ABSTRACT

The emergence of virtual companions is transforming the evolution of intelligent systems that effortlessly cater to the unique requirements of users. These advanced systems not only take into account the user's present capabilities, preferences, and needs but also possess the capability to adapt dynamically to changes in the environment, as well as fluctuations in the user's emotional state or behavior. A virtual companion is an intelligent software or application that offers support, assistance, and companionship across various aspects of users' lives. Various enabling technologies are involved in building virtual companion, among these, Augmented Reality (AR), and Virtual Reality (VR) are emerging as transformative tools. While their potential for use in virtual companions or digital assistants is promising, their applications in these domains remain relatively unexplored. To address this gap, a systematic review was conducted to investigate the applications of VR, AR, and MR immersive technologies in the development of virtual companions. A comprehensive search across PubMed, Scopus, and Google Scholar yielded 28 relevant articles out of a pool of 644. The review revealed that immersive technologies, particularly VR and AR, play a significant role in creating digital assistants, offering a wide range of applications that brings various facilities in the individuals' life in areas such as addressing social isolation, enhancing cognitive abilities and dementia care, facilitating education, and more. Additionally, AR and MR hold potential for enhancing Quality of life (QoL) within the context of virtual companion technology. The findings of this review provide a valuable foundation for further research in this evolving field.

## 1 INTRODUCTION

Virtual companion is revolutionizing the development of intelligent systems that seamlessly adapt to users' specific needs. These advanced systems not only consider the user's current abilities, preferences, and requirements but also have the ability to dynamically adjust to environmental changes and variations in the user's emotional state or behavior. They are made possible through the utilization of cutting-edge companion technologies, which encompass essential features such as competence, individuality, adaptability, availability, cooperativeness, and trustworthiness. These remarkable characteristics are achieved by effectively integrating technical functionality with cognitive processes [1].

Companion technologies are the key enablers components to build the digital or virtual companion. A virtual companion refers to an intelligent and interactive software or application designed to provide support, assistance, and companionship to users in various aspects of their lives. These systems leverage technologies such as artificial intelligence (AI), natural language processing (NLP), machine learning, and Human-Computer Interaction (HCI) to create a personalized and adaptive experience for users [2, 3]. Virtual companion typically embodies human or animal avatars that strive to exhibit realistic and credible behavior. These systems simulate human-like characteristics during face-to-face interactions, encompassing the capacity to perceive and appropriately respond to both verbal and non-verbal cues. They are capable of generating diverse forms of output, including verbal and non-verbal expressions such as mouth, eye, and head movements, hand gestures, facial expressions, and body posture. Furthermore, these systems are adept at handling conversational functions, such as smoothly transitioning between turns, providing feedback, and employing mechanisms for repairing communication breakdowns. The overarching aim is to create virtual companions that closely mirror authentic human behavior and facilitate engaging and natural conversations [4].

In recent years, remarkable advancements in sensor devices and HCI technology have propelled the emergence of a new generation of HCI solutions. This progressive wave of HCI technology places great emphasis on achieving natural interaction and enhanced portability which are essential for virtual companionship. It signifies a shift from earlier HCI approaches reliant on wearable devices towards current methodologies that leverage computer vision, sound, and other non-wearable modalities for HCI. These developments have opened up exciting possibilities for seamless and intuitive interaction, allowing users to engage with technology in a more effortless and unrestricted manner [5]. Among the cutting-edge technologies, immersive technologies such as Augmented Reality (AR), Virtual Reality (VR), and Mixed Reality (MR) play a significant role. These technologies have made substantial contributions to the development of virtual companion. By leveraging AR and VR, immersive environments can be created, allowing users to interact with virtual objects in a manner that feels remarkably real. Current advancements in these technologies aim to utilize advanced AR techniques to enhance the assistance provided to individuals, particularly those facing challenges associated with aging. AR technology provides a real-time view of the physical world, enriched by the addition of virtual computer-generated information. It seamlessly blends real and virtual objects, offering an interactive and three-dimensional experience for users [6].

AR technology holds vast potential across multiple industries, including medicine, the automobile industry, aviation, and space research. Furthermore, its significance extends to the design and development of virtual or digital companions for various applications, such as serving as a personal virtual companion for educational purposes. AR's versatility allows it to enhance medical training, facilitate interactive car or aircraft maintenance guides, assist astronauts in space exploration, and provide immersive educational experiences as a virtual companion [7]. Its adaptability and ability to overlay virtual elements onto the real world make AR a powerful tool for creating engaging and informative interactions in diverse domains. Some examples are virtual culture for humans and pets [8], a simulated virtual companion animal in the form of a dog using AR [9].

VR immersive technology, similar to AR, empowers individuals to access artificial information that transcends the boundaries of a 2D scene. By creating immersive environments, users can engage with unique perceptions that closely resemble the real world. VR technology strives to substitute a person's awareness of the actual environment with an artificial 3D setting. On the other hand, MR technology combines the strengths of VR and AR to deliver an extraordinary and highly immersive encounter that seamlessly blends the digital and physical realms. VR harnesses various human sensory systems, including vision, hearing, touch, and even smell, through output devices, enabling the creation of captivating interactions during user experiences [10]. VR technology offers a myriad of potential applications, spanning across education, medicine, and social interaction. In the realm of virtual companionship, one exciting avenue involves the development of virtual doll companions that facilitate haptic interactions [11]. Additionally, there are efforts to create virtual companions within VR environments that provide multi-sensory feedback, enhancing the immersive experience [12]. These advancements showcase the diverse possibilities of VR technology in fostering interactive and engaging virtual companions, opening up new avenues for meaningful HCI.

The future of the Internet lies in a 3D virtual world where individuals from the physical realm interact with each other and virtual objects using personalized avatars. Within this context of digital companionship, these avatars serve as assistive entities known as virtual agents or, in some instances, virtual humans. They possess the ability to create diverse and intricate scenes within the virtual realm, facilitating natural interactions with objects from the physical world. Central to this virtual environment are immersive technologies that form its primary components. In this virtual world, virtual agents rely on input to initiate specific actions and deliver virtual feedback. Sensors emerge as crucial tools in capturing information from the real environment and relaying it as input to these virtual agents. The integration of immersive technologies like VR, AR, and MR enhances the natural interaction between digital companions and real agents. Consequently, VR, AR, and MR assume vital roles in the development of virtual companions, shaping the future landscape of immersive and intuitive interactions.

Immersive technologies, including VR, AR, and MR, have found applications in diverse fields, such as healthcare [13, 14, 15, 16, 17, 18, 19, 20, 21, 22] and education [23, 24, 25, 26, 27]. Several reviews have explored the potential implications and challenges of these technologies within these domains. However, there is a noticeable gap in systematic research investigating the specific applications of VR, AR, and MR technologies in the development of virtual or digital companion. Consequently, this study was meticulously designed to address this gap by conducting a comprehensive literature review. The primary objective is to explore the contributions of VR, AR, and MR technologies in the development and enhancement of virtual companion. By systematically analyzing the existing body of knowledge, this study aims to shed light on the opportunities and potential areas for improvement within this burgeoning field.

The remainder of this paper is organized as follows: Section 2 outlines the materials and methodology employed for the review, while Section 3 presents the findings derived from the review. Section 4 provides an in-depth discussion and interpretation of the review results, offering insights and analysis. Finally, Section 5 encapsulates the conclusion of this work, summarizing the key outcomes and their implications.

## 2 MATERIAL AND METHOD

This study adhered to the Preferred Reporting for Systematic Reviews and Meta-Analyses (PRISMA) checklist and guidelines, which provide a comprehensive and evidence-based framework for reporting systematic reviews and meta-analyses. By following PRISMA, the authors of this study ensured transparent reporting of the rationale behind conducting the review, the methodologies employed, and the resulting findings. PRISMA serves as a valuable tool in promoting transparency and standardization in the reporting of systematic reviews, enabling readers to accurately assess the quality and validity of the study's methods and outcomes [28].

### 2.1 Goals and Research Questions

The objective of this study is to conduct a comprehensive systematic review that delves into the utilization of immersive technologies (VR, AR, and MR) in the development of digital companions across different domains. In order to encompass a wide range of relevant literature, the following research questions were formulated to guide the investigation:

RQ1: How have immersive technologies (VR, AR, and MR), been applied in the development and utilization of virtual companions across different domains?

RQ2: Which key challenges are identified in the implementation and utilization of virtual companions that leverage immersive environments?

By addressing these research questions, this study aims to provide a comprehensive understanding of the applications and challenges related to the integration of immersive technologies in the realm of digital companionship. The insights gained from this analysis will contribute to the advancement and improvement of future digital companion.

## 2.2 Research Protocol

A meticulous systematic review was undertaken to investigate the utilization of immersive technologies, namely AR, VR, and MR, in the development and utilization of virtual companions. This review diligently adhered to the Preferred Reporting Guidelines for Systematic Reviews and followed the flow chart for Systematic Reviews as outlined in meta-analysis practices as shown in Figure1. By employing these rigorous guidelines, the review ensured methodological rigor, transparency, and replicability in the selection and analysis of relevant literature pertaining to the use of immersive technologies in digital companion systems [28].

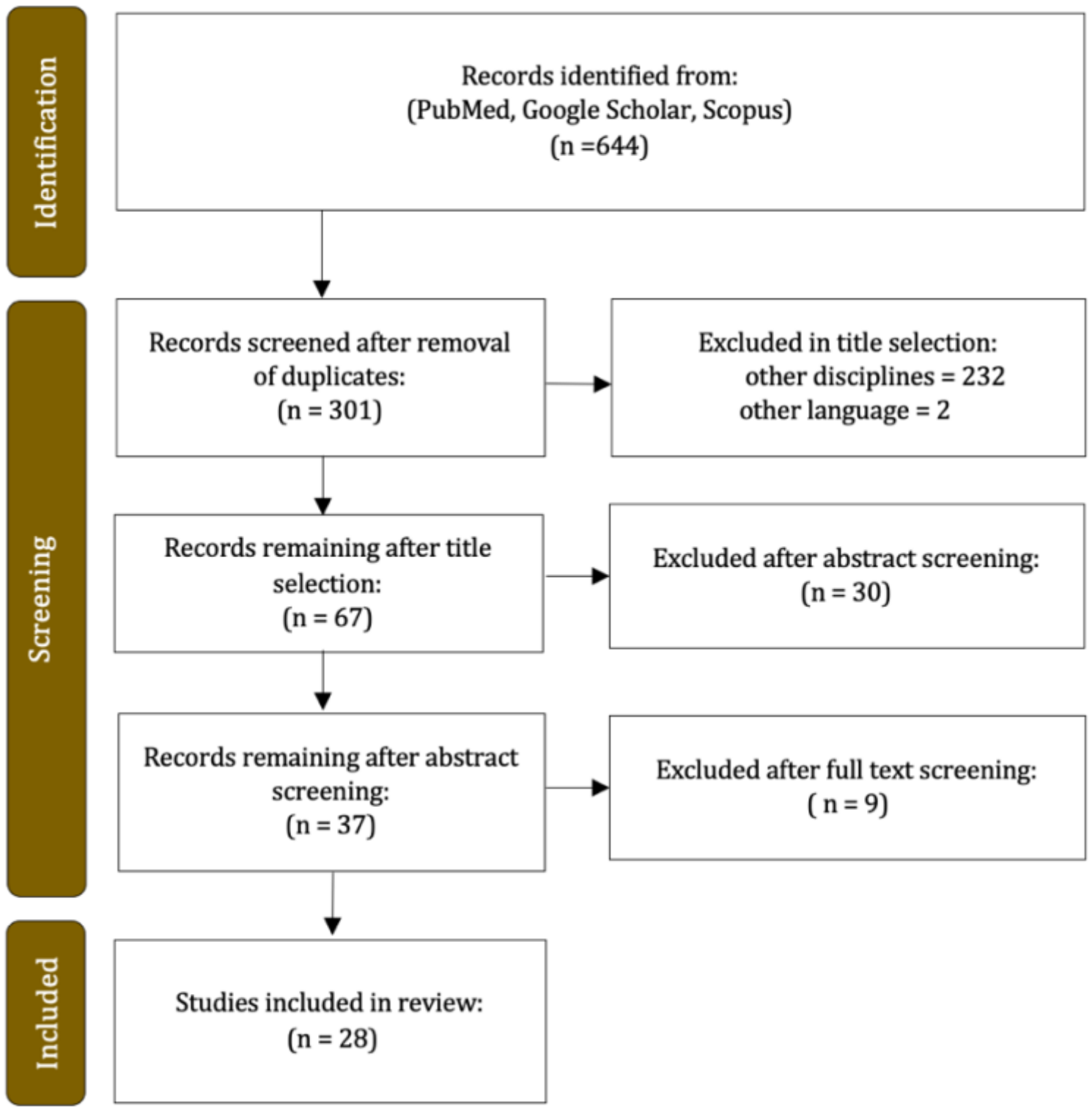

Figure 1: Flowchart illustrating the PRISMA method used for article selection in our study.

## 2.3 Eligibility Criteria

To meet the inclusion criteria for this systematic review, articles had to specifically address the application of immersive technologies (AR, VR, and MR) within the context of digital or virtual companion. Articles published in journals or conference proceedings, encompassing reviews or thematic articles, were considered eligible. Conversely, articles were excluded if they failed to meet the eligibility criteria outlined for this study or if they focused on immersive technologies in contexts unrelated to virtual companion. Following are the exclusion criteria:

Articles that do not meet the eligibility criteria for the study.

Articles focusing on immersive technologies in contexts unrelated to digital companion.

Articles published in non-peer-reviewed sources or sources of low credibility.

Articles not available in the English language.

Articles lacking sufficient information or data to contribute meaningfully to the review.

Articles that primarily focus on hardware or technical aspects of immersive technologies rather than their application in virtual companion.

### 2.4 Search Strategy

This study conducted a comprehensive search of electronic databases, namely PubMed, Scopus, and Google Scholar, to collect relevant articles. The search utilized specific keywords aligned with the concepts explored in the research questions. To enhance the precision of the search, multiple keywords were developed for each concept, as outlined in Table 1. By employing this robust search strategy, the study aimed to obtain accurate and comprehensive results from the databases.

Table 1: Search concepts and keywords used to identify relevant articles for our study.

| Research Question | Concept | Keywords |
|---|---|---|
| RQ1 | Immersive Technology | immersive technologies, virtual reality, augmented reality, mixed reality, assistive technologies |
| RQ1 & RQ2 | Virtual Companion | digital companion, virtual companion, digital assistant, personal assistant |
| RQ1 | Immersive Environment | Immersive environment, 3D environment, virtual world |
| RQ2 | Key challenges | Key challenges, implementation, utilization |

A comprehensive search query was formulated by combining a predefined set of keywords. To ensure a comprehensive and precise result from the databases, various search techniques such as Boolean operators and truncation were employed to effectively combine the keywords.

### 2.5 Characteristics of Included Articles

The inclusion criteria for this review prioritized recent studies conducted between 2016 and 2023, specifically focusing on the utilization of immersive technologies (VR, AR, and MR) for assistive operations. A total of 28 studies were included and thoroughly reviewed, encompassing various aspects of companionship facilitated by VR, AR, and MR. The breakdown of the included studies revealed 1 article from 2023, 4 articles from 2022, 5 articles from 2021, 6 articles 2020, 3 articles from 2019, 2 articles from 2018, 3 articles from 2017, and 4 articles from 2016. These studies explored diverse applications of immersive technologies in the implementation of virtual companions, each serving different purposes. Figure 2 illustrates the percentage distribution of publications across different years.

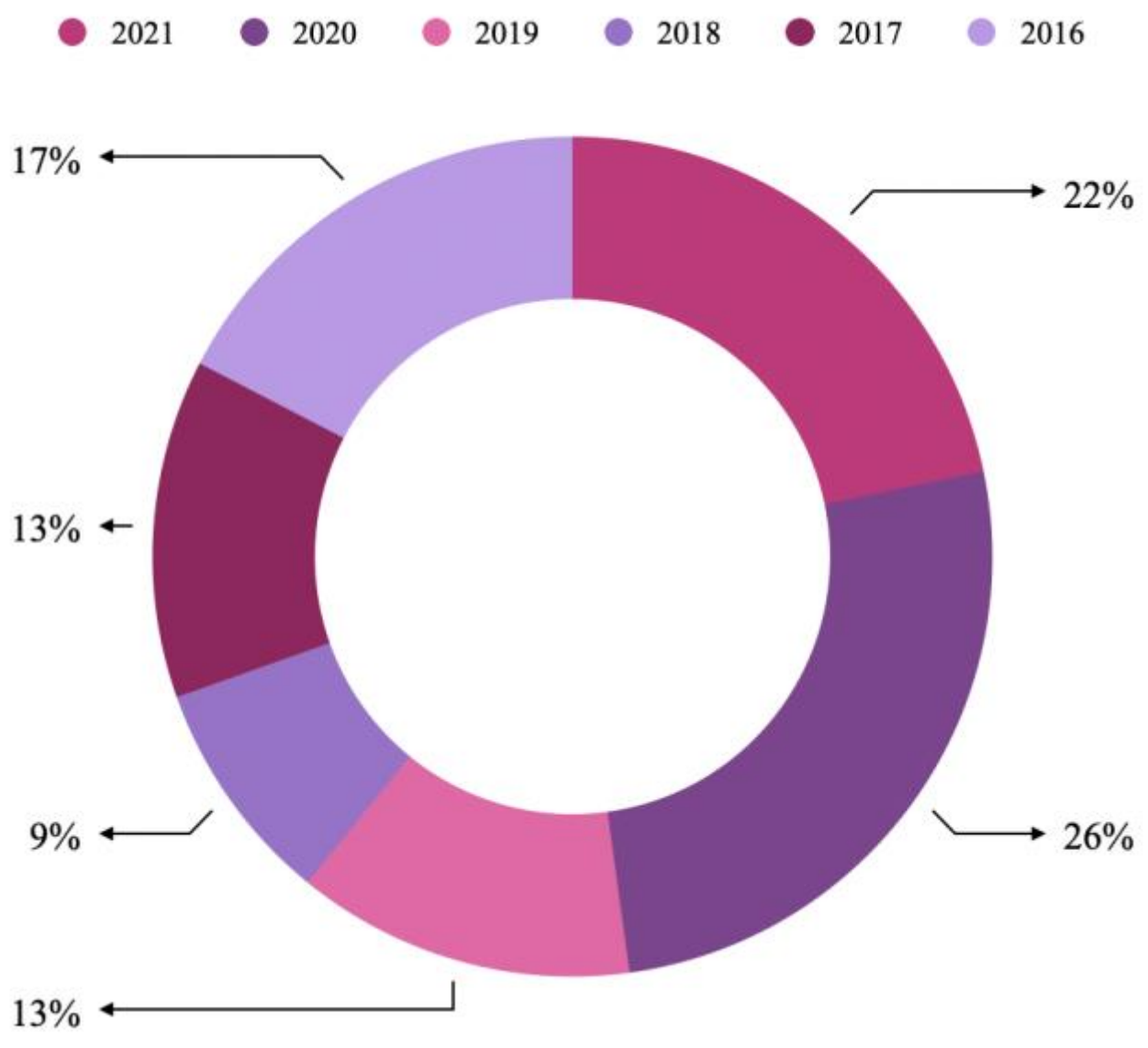

Figure 2: Percentage breakdown of publications across different years

### 2.6 Quality Assessment of the Included Articles

In addition to the specified inclusion and exclusion criteria, the assessment of study quality is an additional criterion employed to determine the significance of each included study. This evaluation involves the use of specific instruments, typically in the form of checklists comprising questions that require thorough assessment. Each quality item within these checklists is assigned numerical scales to aid in the evaluation process. A checklist consisting of 8 criteria was identified from studies [29, 30] to conduct a quality assessment of the included studies that were retained for further analysis. This checklist served as a tool to evaluate the overall quality of the studies and ensure a robust and thorough analysis. Each question within the assessment is rated using a four-point scale. A response of "Yes" is assigned a score of 2 points, "Partially" receives 1 point, "Unable to determine" is assigned a score of 0.5, and a response of "No" is given 0 points. This scoring system enables a more nuanced evaluation of each question's relevance and contribution to the overall quality assessment. Table 2 shows the quality assessment checklist. Table 3 presents the findings of the quality assessment, indicating that all the included articles have met the necessary qualifications and can be considered suitable for further analysis. The quality assessment process has confirmed the reliability and credibility of the selected articles, ensuring their value and relevance for the subsequent analysis.

Table 2: Quality assessment checklist

| No. | Question |
|---|---|
| 1 | Is the objective clearly described? |
| 2 | Is the article well designed the achieve these aims? |
| 3 | Is the study discipline stated clearly? |
| 4 | Are the data collection method(s) described adequately? |
| 5 | Does the study explain the reliability and validity of the measures? |
| 6 | Do the results add to the literature? |
| 7 | Does the study add to your knowledge? |
| 8 | Are the main findings of the study clearly explained? |

Table 3: Quality assessment results

| Article | Q1 | Q2 | Q3 | Q4 | Q5 | Q6 | Q7 | Q8 | Total | Percentage |
|---|---|---|---|---|---|---|---|---|---|---|
| A-1 | 2 | 1 | 2 | 1 | 2 | 2 | 2 | 2 | 14.0 | 87.50% |
| A-2 | 2 | 2 | 2 | 1 | 1 | 2 | 2 | 2 | 14.0 | 87.50% |
| A-3 | 2 | 1 | 1 | 1 | 2 | 2 | 2 | 2 | 13.0 | 81.25% |
| A-4 | 2 | 2 | 2 | 0 | 0 | 2 | 2 | 2 | 12.0 | 75.00% |
| A-5 | 2 | 2 | 2 | 1 | 1 | 2 | 2 | 2 | 14.0 | 87.50% |
| A-6 | 2 | 2 | 2 | 0 | 0 | 2 | 2 | 2 | 12.0 | 75.00% |
| A-7 | 2 | 2 | 2 | 0.5 | 0 | 2 | 2 | 2 | 12.5 | 78.12% |
| A-8 | 2 | 1 | 2 | 0 | 0 | 2 | 2 | 2 | 11.0 | 67.75% |
| A-9 | 2 | 2 | 1 | 0.5 | 0.5 | 2 | 2 | 2 | 12.0 | 75.00% |
| A-10 | 2 | 1 | 2 | 1 | 0 | 2 | 2 | 2 | 12.0 | 75.00% |
| A-11 | 2 | 2 | 2 | 0 | 0 | 2 | 1 | 2 | 11.0 | 67.75% |
| A-12 | 2 | 2 | 2 | 0.5 | 0 | 2 | 2 | 2 | 12.5 | 78.12% |
| A-13 | 2 | 1 | 2 | 1 | 0.5 | 2 | 1 | 2 | 11.5 | 71.87% |
| A-14 | 2 | 2 | 1 | 0 | 0.5 | 2 | 2 | 2 | 11.5 | 71.87% |
| A-15 | 2 | 2 | 2 | 0 | 0 | 2 | 1 | 2 | 11.0 | 67.75% |
| A-16 | 2 | 1 | 2 | 0.5 | 1 | 2 | 2 | 2 | 12.5 | 87.12% |
| A-17 | 2 | 2 | 1 | 0 | 1 | 2 | 2 | 2 | 12.0 | 75.00% |
| A-18 | 2 | 1 | 2 | 1 | 0 | 1 | 2 | 1 | 10.0 | 62.50% |
| A-19 | 2 | 2 | 2 | 0 | 1 | 2 | 2 | 2 | 13.0 | 81.25% |
| A-20 | 2 | 2 | 2 | 2 | 1 | 2 | 2 | 2 | 15.0 | 93.75% |
| A-21 | 2 | 2 | 2 | 1 | 1 | 2 | 2 | 2 | 14.0 | 87.50% |
| A-22 | 2 | 2 | 2 | 2 | 1 | 2 | 2 | 2 | 15.0 | 93.75% |
| A-23 | 2 | 1 | 2 | 1 | 2 | 2 | 2 | 2 | 14.0 | 87.50% |
| A-24 | 2 | 2 | 2 | 2 | 1 | 2 | 2 | 2 | 15.0 | 93.75% |
| A-25 | 2 | 2 | 1 | 1 | 2 | 2 | 2 | 2 | 14.0 | 87.50% |
| A-26 | 2 | 2 | 2 | 2 | 1 | 0.5 | 2 | 2 | 13.5 | 84.37% |
| A-27 | 2 | 2 | 2 | 2 | 0.5 | 2 | 2 | 2 | 14.5 | 90.63% |
| A-28 | 2 | 2 | 2 | 1 | 1 | 2 | 2 | 2 | 14.0 | 87.50% |

## 2.7 Keyword Mapping

In order to better understand the patterns and trends of our literature, assess the impact of the included publications, and identify relevant studies and key concepts to include in our review, keyword mapping was conducted using VOSViewer [31] to generate a co-occurrence map of selected publications, showcasing the relationships between

keywords. Figure 3 illustrates this map, where main keywords are depicted as circles accompanied by labels. The colors represent keyword clusters, and the size of each circle indicates the frequency of occurrences. The proximity of keywords on the map reflects their degree of association, with a larger distance indicating a weaker correlation and a smaller distance indicating a stronger correlation. This visual representation provides insights into the interconnectedness and relationships among the keywords extracted from the selected publications. Table 4 displays the most frequented keywords and their respective linking strength.

In Figure 3, multiple distinct clusters are observed on the map, with notable clusters including virtual reality, augmented reality, and mixed reality. These clusters signify specific research areas within the study of virtual companions. Notably, the nodes representing VR and AR are larger compared to other clusters, indicating their higher frequency and prevalence in the literature. This suggests that VR and AR are prominent topics within the research area of virtual companions. Within the VR cluster, we noticed that eldercare, exercise, depression, sensors, e-learning, and avatars are closely positioned, indicating their semantic similarity and frequent co-occurrence in the literature. Similarly, within the AR cluster, keywords such as companion, visualization, exergame, older adults, and virtual environment are closely positioned, indicating a close relationship and frequent co-occurrence in the literature. This suggests a strong association between these concepts within the context of AR-based virtual companions. Furthermore, within the MR cluster, keywords like digital retail and omnichannel retail are in close proximity, suggesting a tight connection between these concepts. This proximity suggests that MR technology is predominantly utilized in the domain of virtual shopping assistants, as reported in studies [50, 69, 70, 71].

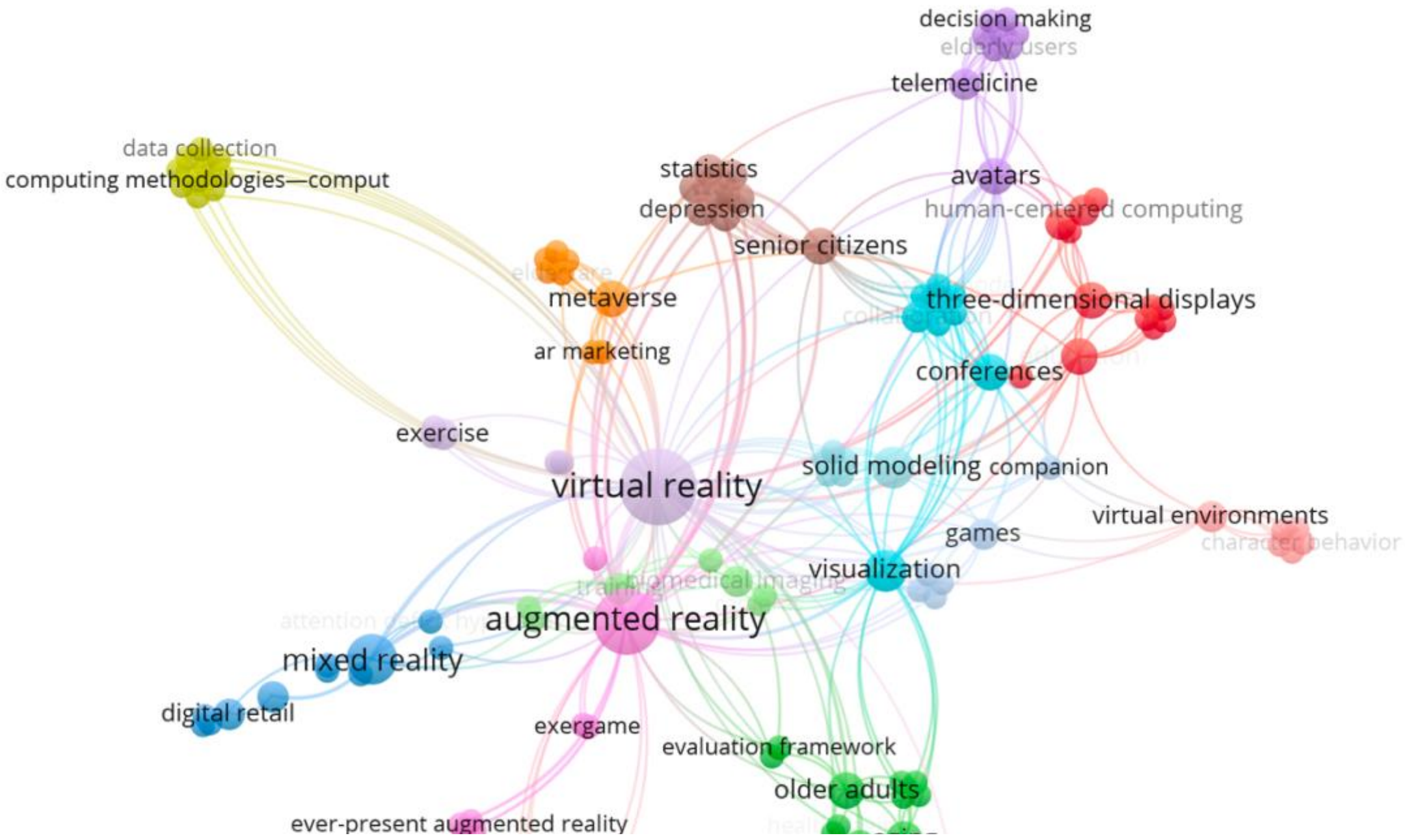

Figure:3 Visualization of Keyword Co-occurrence Network

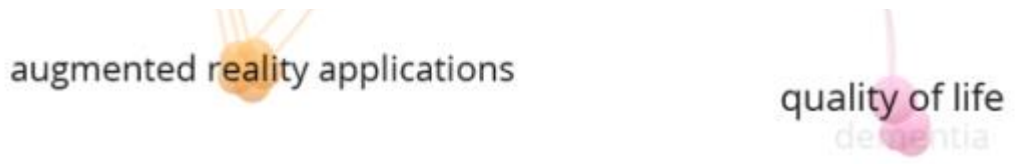

Figure 3: visualization of keyword co-occurrence network

The findings of the keyword mapping analysis revealed distinct research areas within virtual companions, with VR and AR emerging as prominent topics, while also identifying close relationships between keywords related to eldercare, exercise, depression, e-learning, avatars, companionship, visualization, exergame, and virtual environments in the context of immersive technologies.

Table 4: The most co-occurred keyword

| Id | Keyword | Occurrence | Link Strength | Id | Keyword | Occurrence | Link Strength |
|---|---|---|---|---|---|---|---|
| 1 | Virtual Reality | 30 | 78 | 14 | Companion | 18 | 34 |
| 2 | Augmented Reality | 28 | 62 | 15 | Telemedicine | 18 | 32 |
| 3 | Mixed Reality | 27 | 17 | 16 | Companion animals | 16 | 43 |
| 4 | Visualization | 27 | 27 | 17 | Ageing | 13 | 32 |
| 5 | Senior Citizens | 26 | 26 | 18 | Immersive technology | 12 | 33 |
| 6 | Avatars | 26 | 23 | 19 | Social Isolation | 12 | 35 |
| 7 | Education | 26 | 16 | 20 | Exercise | 11 | 24 |
| 8 | Solid Modeling | 25 | 19 | 21 | Quality of Life | 11 | 26 |
| 9 | shopping assistant | 24 | 14 | 22 | Training | 10 | 54 |
| 10 | Virtual pet | 24 | 19 | 23 | 3D displays | 9 | 52 |
| 11 | Virtual environment | 23 | 35 | 24 | Omnichannel retail | 8 | 32 |
| 12 | Social companion | 23 | 43 | 25 | Social interaction | 8 | 16 |
| 13 | Virtual agent | 21 | 34 | 26 | Virtual humans | 7 | 43 |

## 3 RESULTS

We identified a collection of 644 articles published in various journals between 2016 and 2023. Through a meticulous process involving the examination of titles, abstracts, and full texts, we applied our inclusion and exclusion criteria, resulting in a refined selection of 28 publications. Among the remaining articles, the predominant focus was on VR, with 15 articles dedicated to this subject. AR constituted the subject matter of 6 articles, while Mixed Reality was the main topic of 7 articles.

Through our comprehensive review, we examined different applications within each technology category, as illustrated in Figure 4. Our analysis revealed that among the 28 articles, 52% of the articles discussed the utilization of VR in digital companions. Specifically, VR was predominantly employed to enhance the quality of QoL individuals through a diverse range of activities. Within the subset of articles focusing on VR-based digital companions, 6 articles specifically addressed the use of VR to alleviate social isolation among the elderly. Additionally, 4 studies explored the role of VR in enhancing the cognitive abilities of older individuals diagnosed with dementia. The remaining studies explored topics such as fall prevention, companionship during meals, virtual human agents aiding in healthcare, distance learning, and child education. Collectively, these studies indicate that VR holds promise as a valuable tool for improving cognitive function in older adults with dementia. Nonetheless, it is crucial to recognize that further research is necessary to gain a comprehensive understanding of the efficacy and long-term effects of VR on cognition within this population.

AR holds significant potential in the advancement of virtual companions, particularly when combined with other assistive technologies. Its closer proximity to the real world compared to other forms of virtual reality renders AR highly suitable for creating lifelike virtual companions. Our review identified 7 studies that specifically focused on the utilization of AR in digital companions. Among these studies, four examined systems designed to aid elderly individuals, addressing the prevention of social isolation. Another study explored the application of AR in developing digital assistants for assembly tasks within the workplace. Additionally, one study delved into the creation of a digital shopping assistant that leverages AR technology to offer customers a more immersive shopping experience. Lastly, a study centered on providing companionship for cognitive enhancement. Seven studies examined the incorporation of MR, combining VR and AR technologies, on virtual companions. One focused on creating a workplace digital assistant, another aided cognitive improvement for elderly people with dementia, and the third proposed a personal medical assistant, and the remained 4 articles discussed the MR-based digital shopping assistant for individuals.

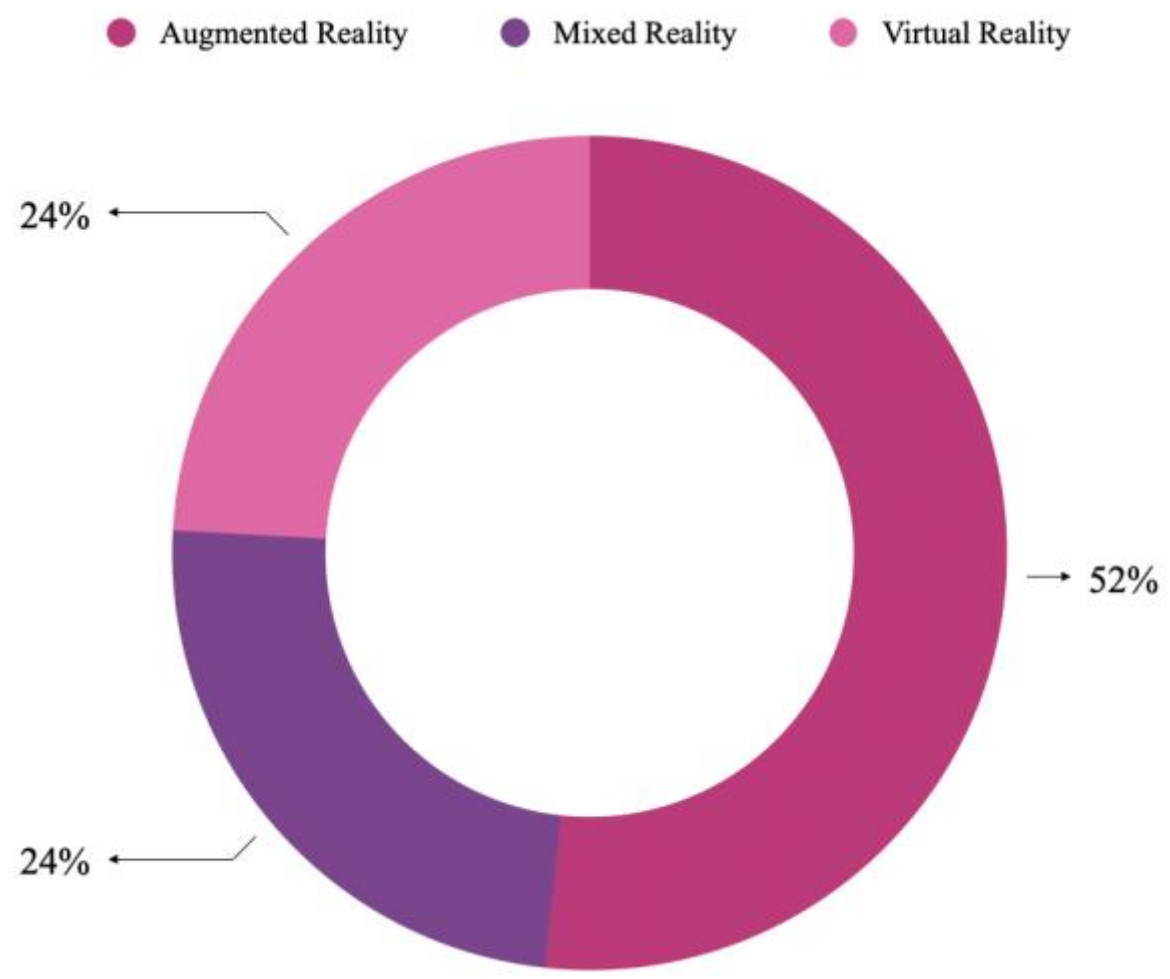

Figure 4: Percentage of VR, AR, and MR articles included in the review.

The literature suggests that VR is an effective tool for immersive virtual companions in elderly care, education, and healthcare. Additionally, AR and MR show promise in assistive applications, including elderly care, cognitive enhancement, education, and addressing glossophobia. Figure 5 depicts the number of applications of three immersive technologies (VR, AR, and MR) in virtual companions, as identified by the review. Generally, this study identified eight distinct categories of VR, AR, and MR applications in virtual companion. These categories encompass the following areas: personal assistant for education, personal assistant for medicine, virtual companion for addressing social isolation, virtual companion for enhancing cognitive abilities, virtual shopping assistant, virtual working environment assistants, virtual companion for the elderly with dementia, and virtual assistant for stress reduction. To provide a clear overview of the findings, Table 5 presents a summary of the reviewed articles in this study, along with a brief explanation of their respective objectives. This table serves as a valuable reference point to understand the key aspects and objectives of each article.

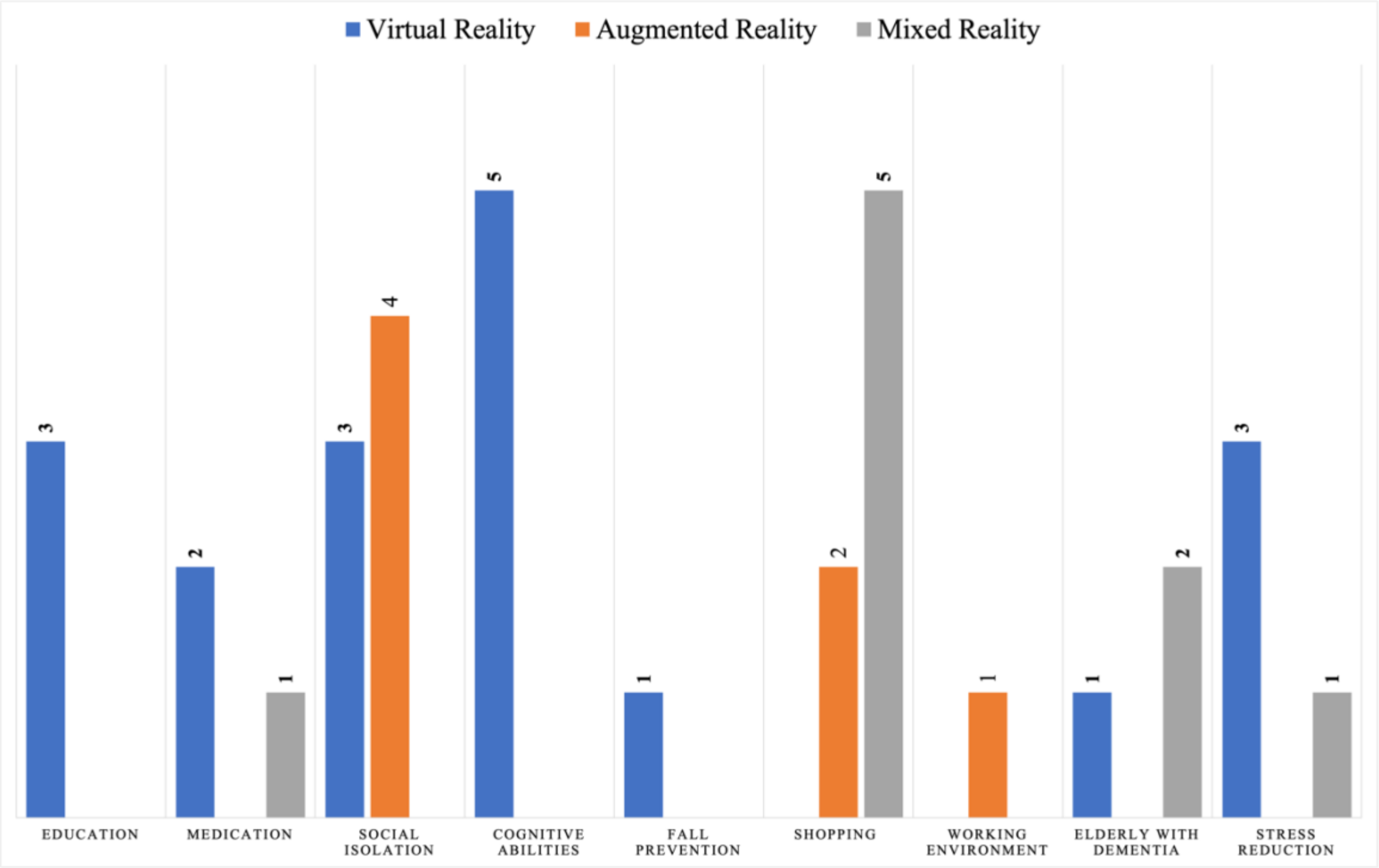

Figure 5: Applications of VR, AR, and MR in Virtual Companions

Table 5: Summary of reviewed articles

| Ref | Year | Technology | Main Objective |
|---|---|---|---|
| [32] | 2017 | VR | Explore the applications of virtual reality in supporting the aging population. |
| [33] | 2016 | VR, AR | An immersive online environment to reduce social isolation and loneliness in the lives of elderly people. |
| [34] | 2016 | VR | A system was developed to assist in identifying difficulties with navigation in individuals with cognitive impairments. |
| [35] | 2018 | VR | The effectiveness of a virtual reality system called ”Virtual Forest” was studied in terms of its impact on engagement, apathy, and mood in people with dementia. |
| [36] | 2016 | MR | A cognitive assistant system was put in place to help older adults with dementia improve their cognitive abilities. |
| [37] | 2020 | MR | A digital assistance system was proposed for use in medicine to provide relevant and contextual information to physicians responsible for time-sensitive pathologies |
| [38] | 2021 | MR | An expert digital assistant was implemented for use in a mixed reality working environment. |
| [39] | 2021 | VR | Prototypes of digital companions were developed for elderly people. These companions were provided through a virtual world in an immersive environment. |
| [40] | 2022 | VR | A virtual reality simulator was developed to aid elderly people in avoiding falls, using a robot navigator and digital twins to facilitate human-robot interaction. |
| [41] | 2017 | VR | Key features were proposed for designing and evaluating assistive applications using avatars and virtual agents for elderly users in the field of assistive technology. |
| [42] | 2018 | VR | The impact of virtual reality on the well-being and social isolation of elderly individuals was studied. |
| [43] | 2019 | VR | Studied virtual presence in distance learning with a virtual companion. |
| [44] | 2016 | VR | A study was conducted on the use of virtual reality to enhance social participation among older people. |
| [45] | 2019 | VR | Implemented system uses virtual reality and a heart rate sensor to help people overcome their speech anxiety and fear of public speaking. |
| [46] | 2020 | VR | The effect of virtual companions in an immersive environment on children's education was studied. |
| [47] | 2022 | AR | A digital assistant using augmented reality was proposed for use in assembly tasks in the workplace. |
| [48] | 2022 | VR | An adaptive virtual assistant was proposed for use in a game environment to aid engineering students in learning manufacturing concepts. |
| [49] | 2021 | AR | A method was proposed for creating a virtual pet using augmented reality technology and the Internet of Things to help elderly people suffering from loneliness and depression. |
| [50] | 2021 | MR | A mixed reality-based digital shopping assistant was developed to provide retail customers with a comprehensive shopping experience. |
| [51] | 2020 | AR | Proposed a virtual dog that act as social companion for elderly and children while they are walking. |
| [52] | 2020 | VR | Proposed an emotional support companion system to reduce the stress of individuals. |
| [53] | 2021 | AR | Developed a virtual pet-based companion system to companion elderly living alone. |
| [54] | 2020 | VR | Developed a virtual birth companion to help to reduce mental stress of the pregnant women during the process of childbirth in labour rooms. |
| [65] | 2020 | AR | AR-based virtual shopping assistant application that helps customers in recommendation and selection of item during shopping. |
| [66] | 2022 | AR | Proposed an AR-based shopping assistant on enhancing the brick-and-mortar store shopping experience through personalized recommendations. |
| [69] | 2023 | MR | Proposed a MR-based personal shopping assistant that using Microsoft HoloLens 2. |
| [70] | 2019 | MR | Developed a MR-based virtual shopping companion capable of recognizing packaged products within retail settings using Microsoft HoloLens. |
| [71] | 2017 | MR | Developed a MR-based shopping assistant application. |

## 4 DISCUSSION

According to this review, immersive technologies such as VR, AR, and MR hold great potential as assistive tools for real-world populations through the provision of virtual companionship in the virtual world. These emerging technologies offer versatile applications for delivering virtual companionship tailored to individual needs and preferences. As these technologies continue to advance and gain wider acceptance, their utilization in virtual companionship is expected to become increasingly common and significant. Furthermore, the development of smart homes, IoT devices, gesture control, and tracking sensors [55] further contributes to the potential improvement of virtual companions. These advancements offer novel opportunities for interacting with technology and the surrounding environment, thus enriching the overall experience, and facilitating greater convenience and comfort.

The objective of this review was to explore the diverse applications of AR, VR, and MR in virtual or digital companions for individuals. To address this objective, the review followed a structured approach, aiming to offer comprehensive insights to answer research questions defined in this study pertaining to these technologies. At first, we focused to discuss the answer of our first research question:  How have immersive technologies (VR, AR, and MR), been applied in the development and utilization of virtual companions across different domains? in order to provide a thorough understanding of this matter, we gained valuable insights by categorizing the utilization of VR, AR, and MR technologies in virtual companion in different areas.

### 4.1 Virtual Reality Technology

VR technology holds significant promise in helping uses as virtual assistance, particularly for older adults. By leveraging VR to support residential communities, it becomes possible to foster improved social and emotional well-being among older individuals through novel avenues of social interaction and engagement [42]. Social isolation poses a significant challenge for older individuals, leading to detrimental mental health outcomes, including depression and other cognitive impairments. This is primarily because social isolation profoundly affects emotional well-being and diminishes one's sense of purpose. Therefore, it becomes crucial to identify strategies that prevent social isolation and foster social connections among older adults, ensuring their mental health and overall well-being remain intact. One notable VR application that addresses this concern is Second Life [56]. By enabling older adults to establish new social relationships, this platform serves as an invaluable tool for social interaction and engagement, effectively combating social isolation and enhancing mental well-being. To assess the effectiveness of this technology, a usability test was conducted involving male and female older participants. The findings indicated a strong inclination among older adults to forge relationships within the VR environment, potentially yielding significant positive effects [32].

Dementia, a neurodegenerative disease primarily prevalent among older adults, significantly impairs cognitive functions. This condition is marked by a decline in mental abilities, including memory, problem-solving, and decision-making. Given its nature, dementia profoundly affects an individual's daily life, independence, and is frequently accompanied by physical and behavioral changes [33]. Dementia necessitates specialized care due to cognitive symptoms, such as memory loss, navigation difficulties, and social isolation, which profoundly affect daily life and independence. Immersive technologies offer opportunities to improve the quality of life for individuals with dementia. Notably, navigation deficits are commonly observed among those with weakened cognitive abilities, making public safety particularly crucial for the elderly population. By implementing an assistance system that detects navigation deficits in individuals with cognitive impairments, the safety of their navigation can be significantly improved, thereby benefiting elderly individuals [34]. In the same way, a VR-based virtual forest system helps people with dementia with engagement, apathy, and mood state [35].

Digital assistants have proven to be highly effective for learning purposes, especially in distance learning scenarios. They enable students to access information and receive support for their studies, regardless of their physical presence in a traditional classroom or learning environment. Moreover, digital assistants offer personalized and interactive learning experiences that promote student engagement and motivation. A notable study [48] showcased the development of an VR based virtual assistant designed to teach manufacturing concepts to engineering students. Through interactive and game-like learning experiences, this digital assistant, leveraging VR technology, presented information and tasks in a more immersive and captivating manner. As a result, students were better able to comprehend and retain the taught concepts. Another study discovered that the utilization of immersive environments and conversational agents, as virtual companions, can enhance students' sense of presence and provide valuable support in the distance education setting [43].

The versatility of VR technology enables its utilization in various ways to address challenges such as stress reduction, enhancement of cognitive abilities, and alleviation of loneliness. For instance, VR systems have been implemented to reduce mental stress experienced by pregnant women during childbirth in labor rooms [54]. Additionally, VR systems can assist individuals with physical disabilities in navigating their surroundings more effectively, while also aiding those with mental health conditions in managing their symptoms. In conclusion, the integration of VR technology within assistance systems holds immense potential for substantially helping people by equipping them with new tools and opportunities to achieve their goals and lead more fulfilling lives.

### 4.2 Augmented Reality Technology

AR is a technology that seamlessly merges the digital and physical realms by overlaying digital information and visuals onto the real world. This is accomplished through the utilization of computer vision techniques, including object recognition, plane detection, face recognition, and motion tracking. These techniques enable AR to identify and track physical surfaces and objects in the surrounding environment, facilitating the integration of digital content with the real-world context [57, 58].

AR technology is commonly employed in conjunction with VR technology to enrich the digital companionship experience within immersive environments. The combined utilization of AR and VR technologies holds tremendous potential in mitigating social isolation among elderly individuals. By incorporating gamification elements into various contexts, AR and VR technologies effectively engage older people who may otherwise be isolated from social activities. This approach fosters a sense of connection and participation, enhancing their overall well-being and mental health[53].

In the workplace, certain employees may encounter challenges in performing specific tasks due to mental deficits or other difficulties. In such situations, the assistance of a supportive aid becomes necessary to facilitate their job responsibilities. AR technology emerges as a particularly valuable tool in this context, as it enables real-time guidance and support for employees tackling tasks that might otherwise pose difficulties. For instance, an AR system can provide step-by-step instructions or visual cues to assist employees in product assembly, enhancing their capability to complete the task successfully. Furthermore, real-time feedback and corrections can be delivered through the AR system, allowing employees to execute their duties with greater accuracy. A digital assistant incorporating AR technology proves to be an excellent solution, especially for operators facing mental deficiencies during assembly tasks. By monitoring their actions in real-time, it offers invaluable support and ensures a smoother workflow [47].

AR plays a crucial role in reducing loneliness by fostering meaningful social connections and providing immersive experiences. AR technology enables individuals to interact with virtual companions such as virtual dogs [51] and pets [53] objects, participate in shared activities, and engage in virtual social environments. A virtual dog companions both elderly individuals and children during their walks, providing them with companionship and alleviating feelings of loneliness. By utilizing AR technology, the virtual pet offers support to the elderly, enabling them to combat loneliness through engaging in various activities. This interactive experience helps keep the elderly occupied and provides them with a sense of purpose and companionship during their walks.

Moreover, digital retail shopping assistants using AR are transforming the online shopping experience. By integrating AR technology, these assistants allow customers to virtually try on clothing, visualize furniture, and test cosmetics before purchase. This seamless and personalized approach bridges the gap between online and offline retail, bringing convenience and interactivity to the fingertips of consumers. With AR-powered shopping assistants, customers can make informed decisions and enjoy a more engaging online shopping journey. In [65], the authors examined the influence of an AR-based virtual shopping assistant application on customer attitude, with a focus on the moderating effect of demographics on customers' inclination to shop in-store. The study involved the development of an application and the participation of users to assess how demographics affected their adoption of the AR-based shopping assistant. The findings indicated that factors such as age, gender, income, and shopping preferences played a role, with younger customers demonstrating a higher propensity for frequent shopping using the assistant. Understanding the impact of demographics on customer attitudes towards shopping assistants enables retailers to determine the suitability of such tools for engaging with their target audience and whether they align with the company's overall strategy. In their recent study, the authors explored the influence of an AR-based shopping assistant on enhancing the brick-and-mortar store shopping experience through personalized recommendations. The findings revealed a positive impact of the AR shopping assistant application on customers' perception of brick-and-mortar shopping experiences. Additionally, the study identified potential areas for improvement of the artifact [66], providing valuable empirical insights for further enhancements.

As result, AR technology is of utmost importance in virtual companions as it brings immersive and interactive experiences, fostering companionship and reducing feelings of loneliness by providing individuals with engaging and lifelike virtual interactions.

### 4.3 Mixed Reality Technology

MR, an integration of AR and VR technologies, stands as a pivotal immersive technology [59]. It has proven instrumental in the development of digital companions across diverse domains, including education, medicine, and supporting relationships with dementia patients. By harnessing the combined potential of AR and VR, MR facilitates the creation of unparalleled and captivating experiences. Notably, MR technology is utilized in digital assistants to support workers in their respective work environments, providing guidance through upcoming tasks, with significant applications in the medical field [38].

Extensive research has been conducted on utilizing MR as a supportive application for individuals with dementia, yielding promising results. The findings highlight the efficacy of an MR-based virtual companion as an effective tool in addressing the challenges associated with dementia in elderly individuals [36]. Within the healthcare sector, there is a growing demand for increased autonomy and decision-making power in care plans and delivery. Digital assistants emerge as valuable assets in fulfilling this need. Specifically, a personalized assistant application utilizing MR technology has been developed to support individuals within the medical field. Tailored for personal use in medicine, this application aims to equip physicians dealing with time-sensitive pathologies with timely, contextually relevant information to aid in their decision-making process [37].

MR technology also plays a crucial role in shopping assistants, serving as a powerful tool for retailers to effectively address customers' needs and compete in the omnichannel retail landscape. Leveraging the capabilities of data and analytics, MR-based shopping assistants offer an optimal solution to enhance the customer experience. As a result, these digital assistants have been recommended for implementation in omnichannel retail environments, aiming to elevate the overall shopping experience [67, 68]. For instance, in [50] a MR-based digital shopping assistant was developed to provide retail customers with a comprehensive shopping experience. In another study the authors developed a MR-based personal shopping assistant that leveraged Microsoft HoloLens 2 as archetypes. The system includes various shopping assistant elements such as product information, reviews, recommendations, product availability, a virtual card, and so on [69]. In a similar vein, Fuchs et al [70] utilized the MS HoloLens to create a MR-based virtual shopping companion capable of recognizing packaged products within retail settings. Likewise, Cheng et al [71] developed a MR-based shopping assistant application that incorporated continuous context awareness and natural interaction techniques, seamlessly integrating digital information into the user interface.

To conclude the findings, indicate that VR and AR are widely recognized as the dominant methods of interaction in immersive environments. VR creates a fully simulated digital environment that users can immerse themselves in, while AR overlays digital information onto the real world. Virtual companions make use of VR and AR technologies to provide interactive experiences to users. These technologies enable users to engage with virtual companions in a dynamic and immersive manner, enhancing the overall interaction and sense of presence. The review highlights the effectiveness of VR and AR in improving the interactivity of virtual companions. By leveraging VR and AR, these systems can offer users a more immersive and engaging experience, fostering a deeper connection with their virtual companions. The interactive nature of VR and AR technologies enables users to have meaningful interactions, promoting a sense of companionship and enhancing the overall effectiveness and impact of virtual companion systems.

The following section delves into the challenges linked to the utilization of immersive technologies such as VR, AR, and MR in virtual companions. This section provides insights to address our second research question.

### 4.4 Challenges

Despite notable progress in the development of virtual companion utilizing immersive technologies, there remain challenges to overcome when implementing VR and AR in digital companionship. Several studies have identified concerns regarding the use of VR headsets, which can impede the adoption of these tools for companionship and assistance, especially in situations involving social isolation and well-being. The unfamiliarity with VR technology can potentially engender a sense of burden, discouraging users from actively embracing its use. This, in turn, may lead to reluctance in using VR, making it challenging for the elderly to engage in social interactions within a virtual environment [60]. The acceptance and preference for virtual agents as digital assistants using immersive technologies pose a particular challenge, especially among specific groups such as the elderly. A study conducted to evaluate the acceptance and effectiveness of interaction among the elderly using a daily digital assistant reveals that this issue remains an open question, warranting further investigation and exploration [61, 62].

The utilization of immersive technologies for virtual assistants presents several challenges, primarily encompassing usability, interaction types, and development costs. Of utmost significance is the design aspect, as it directly influences the psychological well-being of the elderly. Additionally, the creation of VR content necessitates active collaboration and participation from professionals in various fields, thereby potentially incurring substantial costs [63].

The success of VR, AR, and MR experiences relies heavily on immersion, particularly in virtual companion where establishing a sense of presence and connection between the user and the virtual character is paramount. Achieving this requires realistic visual and audio feedback, haptic feedback, and sophisticated AI algorithms to simulate human-like behavior. Immersive technologies play a pivotal role in creating a convincing sense of presence in a virtual space, which should also be interactive to ensure user engagement. However, ensuring the believability of the virtual environment, particularly for the elderly and individuals with dementia, can pose challenges that need to be addressed effectively[64].

The design and physicality of the virtual agent is challenging issue often. In most cases when the virtual animal provides companionship using VR and AR for a specific group of population such as children. For instance, a study realized a child's fear of blame and feeling incompetency while using their virtual dog as walking companion for children [51]. Therefore, conducting further research is essential to address the challenges associated with tools, design considerations, costs, and the unique requirements of each target population when developing virtual companion systems utilizing VR, AR, and MR technologies. By dedicating efforts to explore these areas, we can enhance the development and implementation of virtual companion systems, ensuring they effectively meet the specific needs and preferences of diverse user groups.

## 5 CONCLUSION

Immersive technologies like VR, AR, and MR offer immense potential in enhancing individuals' well-being, combating social isolation, facilitating education, and more. These technologies provide a captivating glimpse into the physical world and find applications across diverse domains such as medicine, education, manufacturing, gaming, and beyond. One notable application involves utilizing these technologies to create digital assistants that aid individuals in areas such as education, productivity, mental health, and other domains. In this study, the applications of VR, AR, and MR immersive technologies in the utilization of virtual companions and its challenges in diverse areas were discussed. To achieve this, a systematic literature review was conducted following the PRISMA guideline. A comprehensive search yielded a total of 28 relevant articles, which were included and thoroughly reviewed. The findings reveals that the VR technology has significant potential for creating virtual companions in virtual environments, while AR and MR technologies also show promise in virtual companionship. It is important to acknowledge that this study primarily concentrated on VR, AR, and MR technologies, with limited exploration of voice assistants and extended reality immersive technology. Furthermore, the scope of this study was specifically centered on the application of VR, AR, and MR immersive technologies in virtual companionship. Future research endeavors will aim to conduct a more comprehensive review, encompassing a broader range of immersive technologies, to deepen our understanding of their potential impact on improving quality of life. This will contribute to the advancement of knowledge in the field and provide valuable insights for enhancing virtual companionship experiences.

## CONFLICT OF INTEREST

On behalf of all authors, the corresponding author states that there is no conflict of interest.

## REFERENCES

[1] Biundo, Susanne, & Andreas Wendemuth. Companion Technology: A Paradigm Shift in Human-Technology Interaction. Springer, 2017.

[2] Jegundo, A. L., Dantas, C., Quintas, J., Dutra, J., Almeida, A. L., Caravau, H., Rosa, A. F., Martins, A. I., Queirós, A., & Rocha, N. P. (2019). Usability Evaluation of a Virtual Assistive Companion. In Á. Rocha, H. Adeli, L. P. Reis, & S. Costanzo (Eds.), New Knowledge in Information Systems and Technologies (pp. 706–715). Springer International Publishing. https://doi.org/10.1007/978-3-030-16184-2_67

[3] SVernon, D., Metta, G., & Sandini, G. (2007). A Survey of Artificial Cognitive Systems: Implications for the Autonomous Development of Mental Capabilities in Computational Agents. IEEE Transactions on Evolutionary Computation, 11(2), 151–180. https://doi.org/10.1109/TEVC.2006. 890274.

[4] Cassell, J. (2000). Embodied conversational interface agents. Communications of the ACM, 43(4), 70–78. https://doi.org/10.1145/332051.332075

[5] Jaimes, A., & Sebe, N. (2007). Multimodal human–computer interaction: A survey. Computer Vision and Image Understanding, 108(1), 116–134. https://doi.org/10.1016/j.cviu.2006.10.019

[6] Saxen, F., Ko¨psel, A., Adler, S., Mecke, R., Al-Hamadi, A., Tu¨mler, J., & Huckauf, A, 2017. Investigation of an augmented reality-based machine operator assistance-system. In Companion Technology (pp. 471-483). Springer, Cham.

[7] Zagoranski, S., & Divjak, S. (2003). Use of augmented reality in education. The IEEE Region 8 EUROCON 2003. Computer as a Tool., 2, 339–342 vol.2. https://doi.org/10.1109/EURCON.2003.1248213

[8] Nijholt, A, 2022. Towards Social Companions in Augmented Reality: Vision and Challenges. In International Conference on Human-Computer Interaction (pp. 304-319). Springer, Cham.

[9] Norouzi, N., Kim, K., Lee, M., Schubert, R., Erickson, A., Bailenson, J., Bruder, G., & Welch, G. (2019). Walking Your Virtual Dog: Analysis of Awareness and Proxemics with Simulated Support Animals in Augmented Reality. 2019 IEEE International Symposium on Mixed and Augmented Reality (ISMAR), 157–168. https://doi.org/10.1109/ISMAR.2019.000-8

[10] Kim, Y. M., Rhiu, I., & Yun, M. H. (2020). A Systematic Review of a Virtual Reality System from the Perspective of User Experience. International Journal of Human–Computer Interaction, 36(10), 893–910. https: //doi.org/10.1080/10447318.2019.1699746

[11] Lee, J.-T., Rajapakse, R. P. C. J., Hung, Y.-P., & Tokuyama, Y. (2018). The Development of a Virtual Doll Companion for Haptic Interaction. 2018 Nicograph International (NicoInt), 92–92. https://doi.org/10.1109/ NICOINT.2018.00038

[12] Huang, J. Y., Hung, W. H., Hsu, T. Y., Liao, Y. C., & Han, P. H, 2019. Pumping life: Embodied virtual companion for enhancing immersive experience with multisensory feedback. In SIGGRAPH Asia 2019 XR (pp. 34-35).

[13] Goharinejad, S., Goharinejad, S., Hajesmaeel-Gohari, S., & Bahaadinbeigy, K. (2022). The usefulness of virtual, augmented, and mixed reality technologies in the diagnosis and treatment of attention deficit hyperactivity disorder in children: An overview of relevant studies. BMC Psychiatry, 22(1), 4. https://doi.org/10.1186/s12888-021-03632-1

[14] Skalidis, I., Muller, O., & Fournier, S. (2022). CardioVerse: The cardiovascular medicine in the era of Metaverse. Trends in Cardiovascular Medicine. https://doi.org/10.1016/j.tcm.2022.05.004

[15] Southworth, M. K., Silva, J. R., & Silva, J. N. A. (2020). Use of extended realities in cardiology. Trends in Cardiovascular Medicine, 30(3), 143–148. https://doi.org/10.1016/j.tcm.2019.04.005

[16] Duman, S., C ̧elik O ̈zen, D., & Duman, S ̧. B. (2022). Metaverse in paedi atric dentistry. European Archives of Paediatric Dentistry, 23(4), 655–656. https://doi.org/10.1007/s40368-022-00733-7

[17] Werner, H., Ribeiro, G., Arcoverde, V., Lopes, J., & Velho, L, 2022. The use of metaverse in fetal medicine and gynecology. European Journal of Radiology, 150.

[18] Tan, T. F., Li, Y., Lim, J. S., Gunasekeran, D. V., Teo, Z. L., Ng, W. Y., & Ting, D. S, 2022. Metaverse and virtual health care in ophthalmology: Opportunities and challenges. The Asia-Pacific Journal of Ophthalmology, 11(3), 237-246.

[19] Ramesh, P. V., Joshua, T., Ray, P., Devadas, A. K., Raj, P. M., Ramesh, S. V., Ramesh, M. K., & Rajasekaran, R. (2022). Holographic elysium of a 4D ophthalmic anatomical and pathological metaverse with extended reality/mixed reality. Indian Journal of Ophthalmology, 70(8), 3116. https: //doi.org/10.4103/ijo.IJO 120 22

[20] Strong, J, 2020. Immersive virtual reality and persons with dementia: a literature review. Journal of gerontological social work, 63(3), 209-226.

[21] Zeng, Y., Zeng, L., Zhang, C., & Cheng, A. S. K. (2022). The metaverse in cancer care: Applications and challenges. Asia-Pacific Journal of Oncology Nursing, 9(12), 100111. https://doi.org/10.1016/j.apjon.2022.100111

[22] Koo, H, 2021. Training in lung cancer surgery through the metaverse, including extended reality, in the smart operating room of Seoul National University Bundang Hospital, Korea. Journal of educational evaluation for health professions, 18.

[23] Barteit, S., Lanfermann, L., B ̈arnighausen, T., Neuhann, F., & Beiers- mann, C. (2021). Augmented, Mixed, and Virtual Reality-Based Head- Mounted Devices for Medical Education: Systematic Review. JMIR Serious Games, 9(3), e29080. https://doi.org/10.2196/29080

[24] Chen, F. Q., Leng, Y. F., Ge, J. F., Wang, D. W., Li, C., Chen, B., & Sun, Z. L, 2020. Effectiveness of virtual reality in nursing education: Meta-analysis. Journal of medical Internet research, 22(9), e18290.

[25] Sankaran, N. K., Nisar, H. J., Zhang, J., Formella, K., Amos, J., Barker, L. T., & Kesavadas, T, 2019. Efficacy study on interactive mixed reality (imr) software with sepsis prevention medical education. In 2019 IEEE Conference on Virtual Reality and 3D User Interfaces (VR) (pp. 664-670). IEEE.

[26] Kov ́acs, P. T., Murray, N., Rozinaj, G., Sulema, Y., & Ryb ́arov ́a, R. (2015). Application of immersive technologies for education: State of the art. 2015 International Conference on Interactive Mobile Communication Technologies and Learning (IMCL), 283–288. https://doi.org/10.1109/ IMCTL.2015.7359604

[27] Mallam, S. C., Nazir, S., & Renganayagalu, S. K. (2019). Rethinking Maritime Education, Training, and Operations in the Digital Era: Applications for Emerging Immersive Technologies. Journal of Marine Science and Engineering, 7(12), Article 12. https://doi.org/10.3390/jmse7120428

[28] Prisma. PRISMA. (n.d.). Retrieved December 2, 2022, from https://www. prisma-statement.org

[29] Al-Qaysi, N., Mohamad-Nordin, N., & Al-Emran, M. (2020). Employing the technology acceptance model in social media: A systematic review. Education and Information Technologies, 25(6), 4961–5002. https://doi.org/10.1007/s10639-020-10197-1

[30] Ma, C. Z.-H., Wong, D. W.-C., Lam, W. K., Wan, A. H.-P., & Lee, W. C.-C. (2016). Balance Improvement Effects of Biofeedback Systems with State-of-the-Art Wearable Sensors: A Systematic Review. Sensors, 16(4), Article 4. https://doi.org/10.3390/s16040434

[31] VOSviewer—Visualizing scientific landscapes. (n.d.). Retrieved June 7, 2023, from https://www.vosviewer.com/

[32] Hughes, S., Warren-Norton, K., Spadafora, P., & Tsotsos, L. E. (2017). Supporting Optimal Aging through the Innovative Use of Virtual Reality Technology. Multimodal Technologies and Interaction, 1(4), Article 4. https://doi.org/10.3390/mti1040023

[33] Goodall, G., Ciobanu, I., Taraldsen, K., Sørgaard, J., Marin, A., Dr ̆aghici, R., Zamfir, M.-V., Berteanu, M., Maetzler, W., & Serrano, J. A. (2019). The Use of Virtual and Immersive Technology in Creating Personal- ized Multisensory Spaces for People Living With Dementia (SENSE- GARDEN): Protocol for a Multisite Before-After Trial. JMIR Research Protocols, 8(9), e14096. https://doi.org/10.2196/14096

[34] Cushman, L. A., Stein, K., & Duffy, C. J. (2008). Detecting naviga- tional deficits in cognitive aging and Alzheimer disease using virtual reality.

Neurology, 71(12), 888–895. https://doi.org/10.1212/01.wnl.0000326262. 67613.fe

[35] Moyle, W., Jones, C., Dwan, T., & Petrovich, T. (2018). Effectiveness of a Virtual Reality Forest on People with Dementia: A Mixed Methods Pilot Study. The Gerontologist, 58(3), 478–487. https://doi.org/10.1093/ geront/gnw270

[36] Sonntag, D, 2015. Kognit: Intelligent cognitive enhancement technology by cognitive models and mixed reality for dementia patients. In AAAI Fall Symposium Series

[37] Croatti, A., Bottazzi, M., & Ricci, A, 2020. Agent-Based Mixed Reality Environments in Healthcare: The Smart Shock Room Project. In Inter- national Conference on Practical Applications of Agents and Multi-Agent Systems (pp. 398-402), Cham.

[38] Spirig, J., Garcia, K., & Mayer, S. (2021). An Expert Digital Companion for Working Environments. Proceedings of the 11th International Conference on the Internet of Things, 25–32. https://doi.org/10.1145/3494322. 3494326

[39] Dai, R., & Pan, Z. (2021). A Virtual Companion Empty-Nest Elderly Dining System Based on Virtual Avatars. 2021 IEEE 7th International Conference on Virtual Reality (ICVR), 446–451. https://doi.org/10.1109/ ICVR51878.2021.9483852

[40] Alves, S. F. R., Uribe-Quevedo, A., Chen, D., Morris, J., & Radmard, S. (2022). Developing a VR Simulator for Robotics Navigation and Human Robot Interactions employing Digital Twins. 2022 IEEE Conference on Virtual Reality and 3D User Interfaces Abstracts and Workshops (VRW), 121–125. https://doi.org/10.1109/VRW55335.2022.00036

[41] Shaked, N. A. (2017). Avatars and virtual agents – relationship interfaces for the elderly. Healthcare Technology Letters, 4(3), 83–87. https://doi. org/10.1049/htl.2017.0009

[42] Lin, C. X., Lee, C., Lally, D., & Coughlin, J. F. (2018). Impact of Virtual Reality (VR) Experience on Older Adults' Well-Being. In J. Zhou & G. Salvendy (Eds.), Human Aspects of IT for the Aged Population. Applications in Health, Assistance, and Entertainment (pp. 89–100). Springer International Publishing. https://doi.org/10.1007/978-3-319-92037-5 8

[43] Krassmann, A. L., Nunes, F. B., Bessa, M., Tarouco, L. M. R., & Bercht, M, 2019. Virtual Companions and 3D Virtual Worlds: Investigating the Sense of Presence in Distance Education. In International Conference on Human-Computer Interaction (pp. 175-192). Springer, Cham.

[44] Baker, S., Waycott, J., Pedell, S., Hoang, T., & Ozanne, E. (2016). Older People and Social Participation: From Touch-Screens to Virtual Realities. Proceedings of the International Symposium on Interactive Technology and Ageing Populations, 34–43. https://doi.org/10.1145/2996267.2996271

[45] Herumurti, D., Yuniarti, A., Rimawan, P., & Yunanto, A. A. (2019). Over- coming Glossophobia Based on Virtual Reality and Heart Rate Sensors. 2019 IEEE International Conference on Industry 4.0, Artificial Intelligence, and Communications Technology (IAICT), 139–144. https://doi.org/10. 1109/ICIAICT.2019.8784846

[46] Thiaville, E., Normand, J.-M., Kenny, J., & Ventresque, A. (2020). Virtual Avatars as Children Companions: For a VR-based Educational Platform: How Should They Look Like? https://hal.science/hal-03068405

[47] Ruffieux, S., Torche, S., Caon, M., & Abou Khaled, O, 2022. Etude exploratoire d'un assistant digital d'aide au montage bas ́e sur la projection en r ́ealit ́e augment ́ee: An Exploratory Study of a Projection-based Augmented Reality Digital Assistant to Facilitate Assembly. In IHM'22: Proceedings of the 33rd Conference on l'Interaction Humain-Machine (pp. 1-12).

[48] Zhao,R.,Zhu,R.,Yang,H.,& Sloan,H.(n.d.).Adaptive Virtual Assistant for Virtual Reality-based Remote Learning.

[49] Cho, M.-G. (2021). A Study on Augmented Reality-based Virtual Pets for the Elderly Living Alone. 2021 International Conference on Information and Communication Technology Convergence (ICTC), 1280–1283. https: //doi.org/10.1109/ICTC52510.2021.9620928

[50] Jain, S., Schweiss, T., Bender, S., & Werth, D. (2021). Omnichannel Retail Customer Experience with Mixed-Reality Shopping Assistant Systems. In G. Bebis, V. Athitsos, T. Yan, M. Lau, F. Li, C. Shi, X. Yuan, C. Mousas, & G. Bruder (Eds.), Advances in Visual Computing (pp. 504–517). Springer International Publishing. https://doi.org/10.1007/978-3-030-90439-5 40

[51] Norouzi, N., Kim, K., Bruder, G., & Welch, G. (2020). Towards Interactive Virtual Dogs as a Pervasive Social Companion in Augmented Reality. ICAT-EGVE 2020 - International Conference on Artificial Reality and Telexistence and Eurographics Symposium on Virtual Environments - Posters and Demos, 2 pages. https://doi.org/10.2312/EGVE.20201283

[52] Graf, L., Abramowski, S., Baßfeld, M., Gerschermann, K., Grießhammer, M., Scholemann, L., & Masuch, M. (2022). Emotional Support Companions in Virtual Reality. 2022 IEEE Conference on Virtual Reality and 3D User Interfaces Abstracts and Workshops (VRW), 634–635. https://doi.org/10.1109/VRW55335.2022.00168

[53] Cho, M.-G. (2021). A Study on Augmented Reality-based Virtual Pets for the Elderly Living Alone. 2021 International Conference on Information and Communication Technology Convergence (ICTC), 1280–1283. https://doi.org/10.1109/ICTC52510.2021.9620928

[54] Sharmila. (n.d.). A virtual birth companion! – Yet another application of virtual reality during the COVID-19 pandemic. Retrieved May 1, 2023, from https://www.jpsiconline.com/article.asp?issn=2214-207X;year=2020;volume=8;issue=2;spage=70;epage=71;aulast=Sharmila

[55] Rahman, Md. A., & Hossain, M. S. (2019). A cloud-based virtual caregiver for elderly people in a cyber physical IoT system. Cluster Computing, 22(1), 2317–2330. https://doi.org/10.1007/s10586-018-1806-y

[56] O'Brien, C. J., Smith, J. L., & Beck, D. E. (2016). Real relationships in a virtual world: Social engagement among older adults in Second Life. Gerontechnology, 15(3), 171–179. https://doi.org/10.4017/gt.2016.15. 3.006.00

[57] Rauschnabel, P. A., Babin, B. J., tom Dieck, M. C., Krey, N., & Jung, T. (2022). What is augmented reality marketing? Its definition, complexity, and future. Journal of Business Research, 142, 1140-1150.

[58] Carmigniani, J., Furht, B., Anisetti, M., Ceravolo, P., Damiani, E., & Ivkovic, M. (2011). Augmented reality technologies, systems and applications. Multimedia Tools and Applications, 51(1), 341–377. https://doi.org/ 10.1007/s11042-010-0660-6

[59] Lee, L. H., Braud, T., Zhou, P., Wang, L., Xu, D., Lin, Z., ... & Hui, P. (2021). All one needs to know about metaverse: A complete survey on technological singularity, virtual ecosystem, and research agenda. arXiv preprint arXiv:2110.05352.

[60] Lee, Li Na, Mi Jeong Kim, & Won Ju Hwang. "Potential of Augmented Reality and Virtual Reality Technologies to Promote Wellbeing in Older Adults." Applied Sciences 9, no. 17 (January 2019): 3556. https://doi.org/ 10.3390/app9173556

[61] Yaghoubzadeh, R., Kramer, M., Pitsch, K., & Kopp, S. (2013, August). Virtual agents as daily assistants for elderly or cognitively impaired people. In International workshop on intelligent virtual agents (pp. 79-91). Springer, Berlin, Heidelberg.

[62] GUIDE Consortium: 'User interaction & application requirements – deliverable D2.1', 2011

[63] Nunes, Fatima L. S., & Rosa M. E. M. Costa. "The Virtual Reality Challenges in the Health Care Area: A Panoramic View." In Proceedings of the 2008 ACM Symposium on Applied Computing, 1312–16. SAC '08. New York, NY, USA: Association for Computing Machinery, 2008. https://doi.org/10.1145/1363686.1363993

[64] Baniasadi, T., Ayyoubzadeh, S. M., & Mohammadzadeh, N. (2020). Challenges and Practical Considerations in Applying Virtual Reality in Medical Education and Treatment. Oman Medical Journal, 35(3), e125. https://doi.org/10.5001/omj.2020.43

[65] Mora, D., Zimmermann, R., Cirqueira, D., Bezbradica, M., Helfert, M., Auinger, A., & Werth, D. (2020). Who Wants to Use an Augmented Reality Shopping Assistant Application? https://doi.org/10.5220/0010214503090318

[66] Zimmermann, R., Mora, D., Cirqueira, D., Helfert, M., Bezbradica, M., Werth, D., Weitzl, W. J., Riedl, R., & Auinger, A. (2022). Enhancing brick-and-mortar store shopping experience with an augmented reality shopping assistant application using personalized recommendations and explainable artificial intelligence. Journal of Research in Interactive Marketing, 17(2), 273–298. https://doi.org/10.1108/JRIM-09-2021-0237

[67] Fuchs, K.; Grundmann, T.; Fleisch, E. Towards Identification of Packaged Products via Computer Vision. In Proceedings of the 9th International Conference on the Internet of Things, Bilbao, Spain, 22–25 October 2019; Association for Computing Machinery: New York, NY, USA, 2019; pp. 1–8.

[68] Willems, K.; Smolders, A.; Brengman, M.; Luyten, K.; Schöning, J. The path-to-purchase is paved with digital opportunities: An inventory of shopper-oriented retail technologies. Technol. Forecast. Soc. Change 2017, 124, 228–242.

[69] Jain, S., Obermeier, G., Auinger, A., Werth, D., & Kiss, G. (2023). Design Principles of a Mixed-Reality Shopping Assistant System in Omnichannel Retail. Applied Sciences, 13(3), Article 3. https://doi.org/10.3390/app13031384

[70] Fuchs, K., Grundmann, T., & Fleisch, E. (2019). Towards Identification of Packaged Products via Computer Vision: Convolutional Neural Networks for Object Detection and Image Classification in Retail Environments. Proceedings of the 9th International Conference on the Internet of Things, 1–8. https://doi.org/10.1145/3365871.3365899

[71] Cheng, K., Nakazawa, M., & Masuko, S. (2017). MR-Shoppingu: Physical Interaction with Augmented Retail Products Using Continuous Context Awareness. In N. Munekata, I. Kunita, & J. Hoshino (Eds.), Entertainment Computing – ICEC 2017 (pp. 452–455). Springer International Publishing. https://doi.org/10.1007/978-3-319-66715-7_61